\documentclass[twocolumn,superscriptaddress,showpacs,preprintnumbers,amsmath
,amssymb,aps,prd]{revtex4}
\usepackage{graphicx}
\usepackage{bm}
\usepackage{color}
\usepackage{hyperref}
\def\bh#1{black hole#1
  (BH#1)\gdef\bh{BH}}
\def\ns#1{neutron star#1
  (NS#1)\gdef\ns{NS}}
\def\bbh#1{binary black hole#1
  (BBH#1)\gdef\bbh{BBH}}
\def\gw#1{gravitational wave#1
  (GW#1)\gdef\gw{GW}}
  \def\jbd#1{Brans-Dicke#1
  (JBD#1)\gdef\jbd{BD}}
  \def\stg#1{scalar-tensor#1
  (ST#1)\gdef\stg{ST}}
\def\gr#1{general relativity#1
  (GR#1)\gdef\gr{GR}}
   
\begin{document}

\title{Late Inspiral and Merger of Binary Black Holes in Scalar-Tensor Theories of Gravity}

\author{James Healy}
\affiliation{Center for Relativistic Astrophysics and
School of Physics\\
Georgia Institute of Technology, Atlanta, GA 30332}
\author{Tanja Bode}
\affiliation{Center for Relativistic Astrophysics and
School of Physics\\
Georgia Institute of Technology, Atlanta, GA 30332}
\author{Roland Haas}
\affiliation{TAPIR, California Institute of Technology, Pasadena, CA 91125}
\author{Enrique Pazos}
\affiliation{Departamento de Matem\'aticas,
Universidad de San Carlos,
Guatemala, Guatemala}
\author{Pablo Laguna}
\affiliation{Center for Relativistic Astrophysics and
School of Physics\\
Georgia Institute of Technology, Atlanta, GA 30332}
\author{Deirdre M. Shoemaker}
\affiliation{Center for Relativistic Astrophysics and
School of Physics\\
Georgia Institute of Technology, Atlanta, GA 30332}
\author{ Nicol\'as Yunes}
\affiliation{Department of Physics, Montana State University, Bozeman, Montana 59717}

\begin{abstract} 
  Gravitational wave observations will probe non-linear gravitational
interactions and thus enable strong tests of Einstein's theory of
general relativity. We present a numerical relativity study of the
late inspiral and merger of binary black holes in scalar-tensor
theories of gravity. We consider black hole binaries in an
inhomogeneous scalar field, specifically binaries inside a scalar
field bubble, in some cases with a potential. We calculate the
emission of dipole radiation. We also show how these configurations
trigger detectable differences between gravitational waves in
scalar-tensor gravity and the corresponding waves in general
relativity. We conclude that, barring an external mechanism to
  induce dynamics in the scalar field, scalar-tensor gravity binary
  black holes alone are not capable of awaking a dormant scalar field,
  and are thus observationally indistinguishable from their general
  relativistic counterparts.
\end{abstract}

\maketitle

%%%%%%%%%%%%%%%%%%%%%%%%%%%%%%%%%%%%%%%%%%%%%%%%%%%%%%%%%%%%%%%%%%%%%%%%%%%%

Experimental tests in connection with gravitational redshift, light
deflection, Shapiro time delay and perihelion advance, among others,
have increased our confidence that \gr{} is the correct theory of
gravity~\cite{will:2006:cbg}. More compelling evidence is provided by
binary pulsar observations~\cite{126}, where the hardening of the
binary is accounted for in exquisite detail by one of the fundamental
predictions of \gr{:} \gw{} emission. Modified gravity theories are
nonetheless a possibility~\cite{will:2006:cbg}. To verify that indeed
the podium only belongs to Einstein's theory, strong-gravity tests are
needed, involving for instance core-collapse supernovae or the last
few orbits and coalescence of compact objects. The new astronomy of
\gw{} observations will soon deliver such opportunities, probing
gravity at its strongest grip, and thus testing whether Einstein was
correct.

In anticipation of \gw{} observations by LIGO, Virgo and other
interferometric detectors, exploring what to expect from alternative
theories is crucial in assisting data analysis efforts. Of particular
importance is the investigation of predictions from alternative
theories of gravity during the {\emph{generation}} of \gw{s} in the
non-linear regime. This calls for theoretical studies only accessible
with the tools of numerical relativity, which is the focus of the
present paper. Among the competing alternatives to \gr{,} \stg{}
gravity~\cite{1992CQGra...9.2093D,Damour:1993hw} is one of the most
popular due to its simplicity, and because of the motivations provided
by string theory scenarios and explanations to dark
energy~\cite{11496}. \stg{} gravity in its simplest form was proposed
about a half century ago and is commonly known as Brans-Dicke
theory~\cite{Brans:1961sx}.

Studies on observational consequences of \stg{} theories have mostly
focused on compact object binaries in the post-Newtonian or
extreme-mass-ratio
regimes~\cite{1989ApJ...346..366W,2002PhRvD..65d2002S,2010PhRvD..81f4008Y,
  2009PhRvD..80d4002S,2005CQGra..22S.943B,Will:1994fb,
  2004CQGra..21.4367W,2011arXiv1110.5950Y,2011PhRvD..84f4016G,2011arXiv1112.3351Y},
a critical prediction of which is the emission of dipole
radiation. The latter depends on the sensitivity, $s$, of the compact
objects~\cite{will:2006:cbg}, which is a measure of how susceptible
the mass of an object is to variations in the local value of the
gravitational constant ($s=1/2$ for \bh{s}, and $s \le 1/5$ for
neutron stars~\cite{1989ApJ...346..366W}). With modifications to \gr{}
entering as the difference $\Delta s = s_1-s_2$ of the binary
components, \bbh{s} are unaffected in \stg{} theories since $s=1/2$
for all \bh{s} regardless of their masses or spins. For initially
stationary scalar fields, Will and Zaglauer~\cite{1989ApJ...346..366W}
proved that \bbh{s} in Brans-Dicke theory are indistinguishable from
binaries in \gr{} to first post-Newtonian order.  Similarly, Yunes,
Pani and Cardoso~\cite{2011arXiv1112.3351Y} recently extended this
proof to all \stg{} theories and to all post-Newtonian orders, but to
leading order in the mass ratio.  If the scalar field is not initially
stationary, however, Horbatsch and Burgess~\cite{2011arXiv1111.4009H}
suggested that \bbh{s} could retain scalar
hair~\cite{1999PhRvL..83.2699J} and emit dipole radiation, provided
the holes have unequal masses.

The goal of the present study is to investigate whether the
conclusions from post-Newtonian and extreme-mass ratio studies
regarding the \bbh{} indistinguishability between \stg{} and \gr{}
theories carries over to the non-linear, comparable-mass regime. Our
results confirm this indistinguishability, unless there exists a
mechanism to induce dynamics in the scalar field.  When the latter
activates, the scalar field dynamics triggers dipole energy loss that
leads to detectable differences in the `$+$' and `$\times$' \gw{}
polarizations.  In the present study, we induce scalar field dynamics
by placing the \bbh{s} inside a scalar field bubble, which in some
cases includes a potential. These inhomogeneities in the scalar field
have a dramatic effect on the \bbh{} dynamics and thus noticeable
differences on the \gw{} emission. We show that the changes on \bbh{}
dynamics are due to accretion of scalar field by the merging \bh{s}.

We restrict attention to \stg{} theories in vacuum with a (Jordan frame) action
of the form:
\begin{equation}
\label{eq:Sjordan}
S = \int \frac{d^4x}{2\, \kappa}\,\sqrt{-\tilde
  g}\left[F(\varphi)\,\widetilde R
  -\zeta(\varphi)\widetilde\nabla_\mu\varphi\widetilde\nabla^\mu\varphi 
  - 2\,U(\varphi)\right]
\end{equation}
with $\kappa = 8\,\pi\,G$. Under a conformal transformation
$g_{\mu\nu} = F(\varphi)\,\tilde g_{\mu\nu}$, or equivalently $\tilde
g_{\mu\nu} = A^2(\phi) g_{\mu\nu}$, the action (in the Einstein
frame) reads~\cite{1992CQGra...9.2093D,Damour:1993hw}:
\begin{equation}
  S =\int d^4x\,\sqrt{-g}\left[R/(2\,\kappa) - 
\nabla_\mu\phi\nabla^\mu\phi/2 - \,V(\phi)\right] 
\label{eq:Seinstein}
\end{equation}
where $(d\phi/d\varphi)^2= \left[(\zeta/F) +
  (3/2)(F_{,\varphi}/F)^2\right]/\kappa$ and $V = U/(\kappa\,F^2)$.
We set $A(\phi)= e^{a\,\phi -
  b\,\phi^2/2}$~\cite{1992CQGra...9.2093D,Damour:1993hw}. Thus,
$F=\varphi$ and $\zeta = \omega/\varphi$ with $\omega =
-3/2+\kappa/(a-b\,\phi)^2$. Brans-Dicke theory is recovered when $b=
0$, and \gr{} when $a = b = 0$. We focus here on the case $a=0$, and
consider different values of $b$.

The Einstein frame is convenient because the action in
(\ref{eq:Seinstein}) yields the same equations as those of \gr{},
namely $ G_{\mu\nu} = \kappa\,T_{\mu\nu}$ with $T_{\mu\nu} =
\nabla_\mu\phi\nabla_\nu\phi -
g_{\mu\nu}\left(\nabla_\rho\phi\nabla^\rho\phi/2 + V\right)$ and $\Box
\phi = V_{,\phi}$. Therefore, in the absence of a potential, $\phi=
\phi_0 =$ constant yields $G_{\mu\nu} = 0$. Thus, vacuum spacetimes in
\stg{} theories will be equivalent to their corresponding spacetimes
in \gr{,} independent of the choice of conformal factor
$A(\phi)$. This is also the case in the presence of a potential if one
arranges for $V(\phi_0) = 0$ and $V_{,\phi}(\phi_0) = 0$. We carried
out simulations of spacetimes containing \bh{} singularities that
verify this exact equivalence between \stg{} and \gr{,} differing only
at the level of round-off errors.

Therefore, the only avenue to trigger differences between \stg{}
theories and \gr{} is by inducing dynamics in $\phi$. One possibility
is with $\partial_t\phi \ne 0$, e.g. $\partial_t\phi = \epsilon\,t
$~\cite{2011arXiv1111.4009H}. Another possibility is by introducing a
direct scalar-field curvature coupling to the action, such as $\phi \,
R_{\mu \nu \delta \rho} R^{\mu \nu \delta
  \rho}$~\cite{2011PhRvD..83j4002Y,2011arXiv1110.5950Y} or $\phi \,
\epsilon^{\alpha \beta}{}_{\delta \rho} R_{\mu \nu \alpha \beta}
R^{\mu \nu \delta
  \rho}$~\cite{2009PhRvD..79h4043Y,2009PhR...480....1A,2011arXiv1110.5950Y},
that anchors the scalar field to the spacetime curvature. Yet another
alternative is by introducing inhomogeneities in $\phi$, which is the
focus of the present study.

We consider \bbh{} inside a scalar field \emph{bubble} with an initial
profile $\phi(r) = \phi_0 \, \tanh{[(r - r_0)/\sigma]}$ and an
inflationary-inspired potential $V = \lambda ( \phi^2 - \phi_0^2
)^2/8$ that yield discrete symmetry breaking~\cite{PhysRevD.39.1558}.
The bubble has radius $r_0$, thickness $\sigma$, and in the exterior
(interior) $\phi = \phi_0\, (-\phi_0)$. Notice that, since we are
interested in asymptotically flat spacetimes,
$\phi(r\rightarrow\infty) = \phi_0$; we shift the conformal factor as
$A=e^{-b\,(\phi^2-\phi_0^2)/2}$, so the Einstein and Jordan frames are
both Minkowskian and in the same
coordinates~\cite{1992CQGra...9.2093D,Damour:1993hw}. This allows us
to make \emph{fair} comparisons between \bbh{s} in \stg{} and \gr{}
theories, i.e.~comparisons among \bbh{s} with the same \bh{} masses,
spins, separation and eccentricity.  If $V\ne 0$, the bubble becomes a
topologically stable, domain wall with $\sigma \approx
2/(\phi_0\sqrt{\lambda})$ and surface energy density $\mu \approx
3\,\phi_0^3\sqrt{\lambda}/4$~\cite{PhysRevD.39.1558}. Since in the
interior $V(-\phi_0) = 0$ and $V_{,\phi}(-\phi_0) = 0$, the spacetime
in the neighborhood of the \bbh{} can be arranged to be locally,
nearly equivalent to its \gr{} counterpart.

\begin{table}[htbp]
    \centering%
    \begin{tabular}{l|ccccc}
      Case & $\phi_0$ &   $4\,\pi\lambda/M^2$ & $M_{0}/M$ & $M_{h}/M$ &   $a/M_{h}$ \\
      \hline
      A & 0       & 0           & 0.990 & 0.952 & 0.686 \\ 
      B  & 1/80 & 0           & 1.179 & 0.963 & 0.688 \\
      C  & 1/40 & 0           & 1.747 & 0.999 & 0.706 \\
      D  & 1/80 & $10^3$ & 1.217 & 0.983 & 0.685 
\end{tabular}
   \caption{Parameters, masses and spin.}
    \label{table1}
\end{table}

We discuss results from a set of representative simulations in our
study. The simulations were obtained with the \textsc{Maya} numerical
relativity code of our group~\cite{Bode:2009mt,2011arXiv1111.3344L},
modified to include the scalar field $\phi$ in the Einstein frame. In
all cases, the binary has non-spinning, equal-mass \bh{s} in
quasi-circular orbit, initially separated by $11\,M$, with $M$ the
mass of the binary. The bubble surrounding the \bbh{} has a radius
$r_0 = 120\,M$ and thickness $\sigma = 8\,M$. The simulations in the
Einstein frame differ only in the parameters $\phi_o$ and $\lambda$,
as given in Table~\ref{table1}. Also in Table~\ref{table1} are the
values in the Einstein frame of the ADM mass $M_0$, the mass $M_h$ and
spin $a/M_h$ of the final \bh{.} Case A is the reference \gr{}
simulation, and D is the only case with non-vanishing
potential. Extraction of \gw{s} is carried out in the Jordan
frame. For each case, we extract \gw{s} with values $b = 0, 5$ and 10
for the parameter in the conformal factor
$A=e^{-b\,(\phi^2-\phi_0^2)/2}$.

\begin{figure}[ht]
\includegraphics[width=0.45\textwidth]{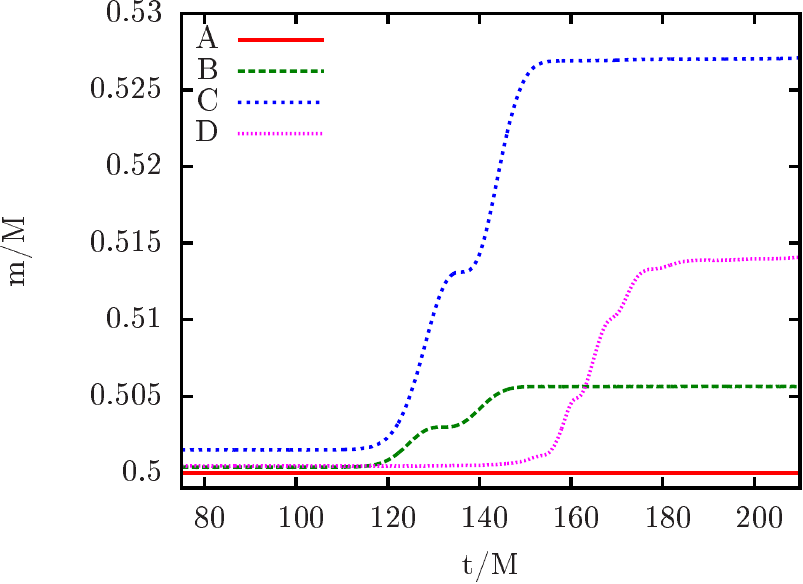}
\caption{Mass of the individual \bh{s} as they accrete $\phi$ before
  they merge. Here $t = 0\,M$ is the initial time of the simulation.}
\label{fig:masses}
\end{figure}

In all cases, the bubble shell collapses. For case D (a domain wall
bubble), the thickness of the shell does not change much during its
collapse.  For cases B and C with $V=0$, however, the shell disperses
while collapsing. By the time the shell reaches the \bbh{}, the cloud
has encompassed the binary. Independent of the details of the initial
configuration, the scalar field has a dramatic effect on the \bbh{}
dynamics. When the bubble reaches the binary, each \bh{} accretes a
significant amount of mass from the scalar field.
Figure~\ref{fig:masses} shows the evolution of the mass (in the
Einstein frame) of the individual \bh{s} before they merge. The
differences in the initial \bh{} masses among all cases reflect the
difficulty of setting up identical binaries.  Notice that for cases B
and C the mass increases in two stages. The first increase occurs when
the bubble is imploding and passes through the binary. The second
increase is after the bubble bounces back and expands. The mass
increase in C is larger because the bubble is more massive. With the
choice of parameters, the most massive shell is C, followed by D and B
(see the ADM mass $M_0$ in Table~\ref{table1}).  For case D, the
increase in the \bh{} mass as a consequence of $\phi$-accretion
differs from cases B and C in the presence of multiple step-like
increases.  This is because in case D the bubble does not just bounce
back and dissipate, but instead it lingers in the neighborhood of the
binary, bouncing multiple times due to the potential and leading to
the series of step increases.  Notice that for all cases, the \bh{s}
eventually reach a constant mass.  The \bh{s} gain 1\%, 6\% and 3\% of
their original mass for cases B, C and D, respectively.  This gain in
mass is correlated with the mass $M_h$ of the final \bh{.}  That is,
in the \gr{} case A, the final \bh{} is $M_h = 0.952\,M$, while in all
other cases and to a good approximation, the final \bh{} mass is $M_h
\approx 0.952\,(M+\delta M)$ with $\delta M$ the mass accreted by the
merging \bh{s}, i.e. $0.01\,M$ for B, $0.06\,M$ for C, and $0.03\,M$
for D (see e.g.~Table~\ref{table1}).
\begin{figure}
\includegraphics[width=0.4\textwidth]{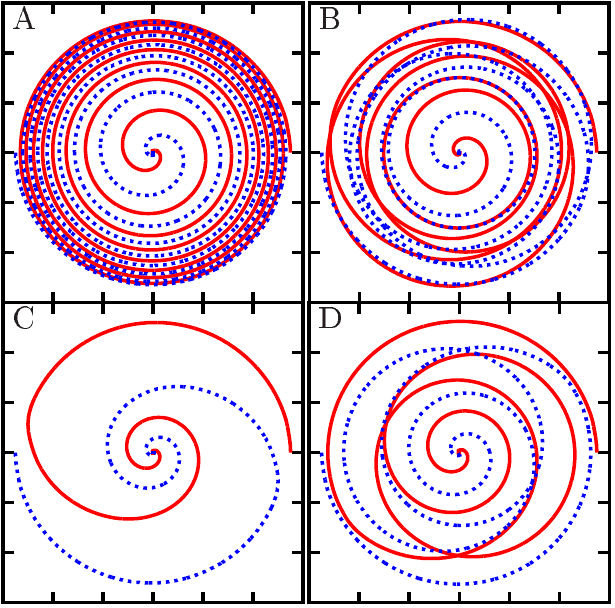}
\caption{Trajectories of the \bh{s} when viewed from the Einstein
  frame.}
\label{fig:trajectories}
\end{figure}

The $\phi$-accretion by the \bh{s} has a dramatic effect on the binary
dynamics.  Figure~\ref{fig:trajectories} shows the trajectories (in
the Einstein frame) for all cases. Recall that case A is the \gr{}
reference case for a \bbh{} in quasi-circular orbit. In all the other
cases, the $\phi$-bubble induces eccentricity and accelerates the
merger. The more massive the shell the larger the induced
eccentricity. In particular, for C the influence of $\phi$ is such
that the binary basically plunges.

\begin{figure}[ht]
\includegraphics[width=0.45\textwidth]{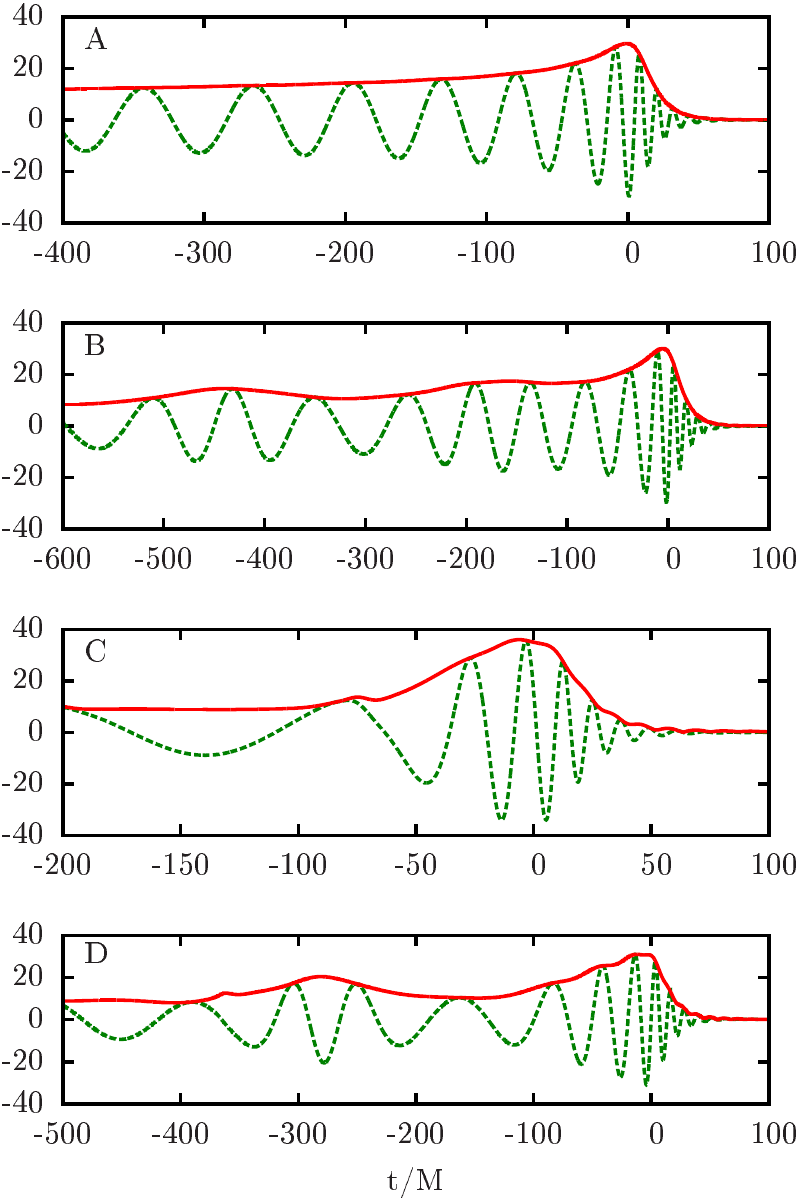}
\caption{Mode (2,2) of the gravitational strain $(r/M)\,h_+$ (dashed
  line) and the amplitude $(r/M)|h| = (r/M)|h_+-i\,h_\times|$ (solid
  line). The time $t = 0\,M$ denotes merger time.}
\label{fig:strain}
\end{figure}

Not surprisingly, the effect of the bubble is also felt in the
emission of gravitational radiation.  In Figure~\ref{fig:strain}, we
plot the (2,2) mode of the \gw{} strain polarization $h_+$ (dashed
line) and the amplitude $|h| = |h_+-i\,h_\times|$ (solid line), with
$t = 0\,M$ denoting merger time.  The strains plotted in panels B, C
and D correspond to $b=0$. Notice the obvious differences between the
strains, and in particular when matched against the \gr{} case
A. However, within each of the B, C and D cases, the differences
between a strain with $b=0$ and one with $b\ne 0$ are
undetectable. That is, the conformal factor does not add new
features. We calculated mis-matches between the $b=0$ and $b\ne 0$
strains for Adv.~LIGO, within each B, C and D case, and found them to
be $\lesssim 10^{-3}$. This can be understood by studying how the Weyl
scalar $\Psi_4 = -C_{l\bar m l \bar m}$ conformally transforms. The
tetrad vectors transform as $\tilde v^a = A^{-1}\, v^a$, and the Weyl
tensor as $\widetilde C_{abcd} = A^2\, C_{abcd}$, and thus
$\widetilde\Psi_4 = A^{-2}\, \Psi_4 =
e^{b\,(\phi^2-\phi_0^2)}\,\Psi_4$ (tilde quantities correspond to the
Jordan frame). During the course of the simulation, we observed that
the scalar field at the location of the \gw{} extraction behaves as
$\phi = \phi_0 + \delta\phi$ with $\delta\phi/\phi_0 \lesssim
10^{-2}$. Therefore, $\widetilde \Psi_4 \approx (1+
2\,b\,\delta\phi\,\phi_0)\,\Psi_4 \approx (1+
10^{-2}\,b\,\phi_0^2)\,\Psi_4$. Since for the chosen parameters
$b\,\phi_0^2 \lesssim 10^{-2}$, $\widetilde \Psi_4 \approx \Psi_4$ to
one part in $10^{4}$.

In Figure~\ref{fig:strain}, cases B and D show the characteristic
strain modulation observed in eccentric \bbh{s}. The waveform in case
C, on the other hand, is basically a burst, since here the binary
essentially plunges. Notice also that the amplitudes in C and D show
small bumps, the first one at $t \sim -75\,M$ for case C, and at $t
\sim -375\,M$ for case D. These bumps are due to the shell as it goes
through the binary. The bumps are therefore correlated with the mass
jumps in Fig.~\ref{fig:masses}.

\begin{figure}[ht]
 \includegraphics[width=0.45\textwidth]{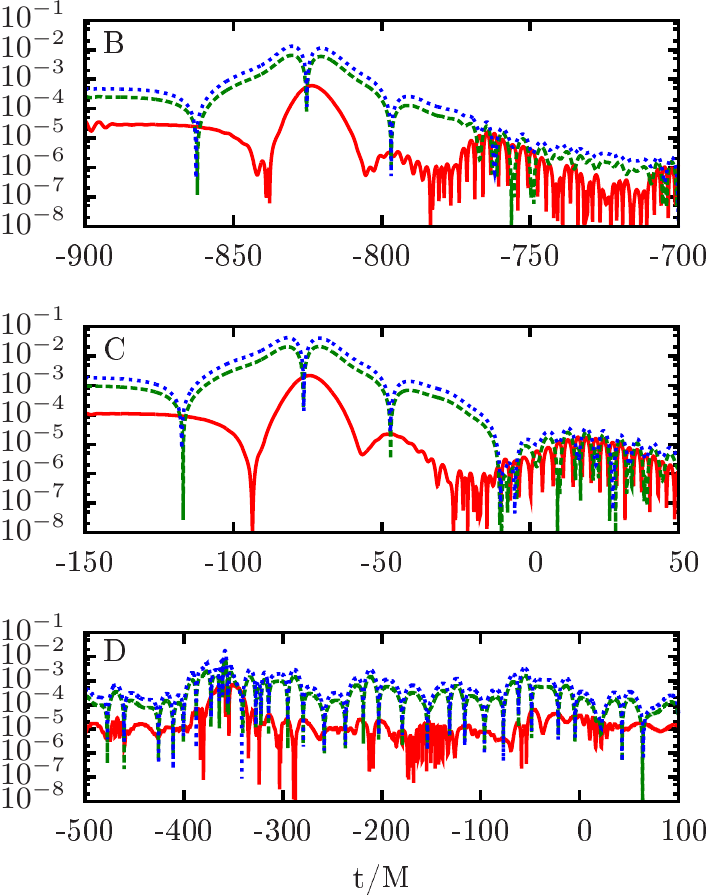}
  \caption{Mode (0,0) of the breathing mode, $r\,M\,\widetilde
    \Phi_{22}$. In each panel, $b=0, 5$ and 10 are solid, dashed, and
    dotted lines, respectively.  $t = 0\,M$ denotes merger time. }
\label{fig:Phi00}
\end{figure}

\begin{figure}[ht]
  \includegraphics[width=0.45\textwidth]{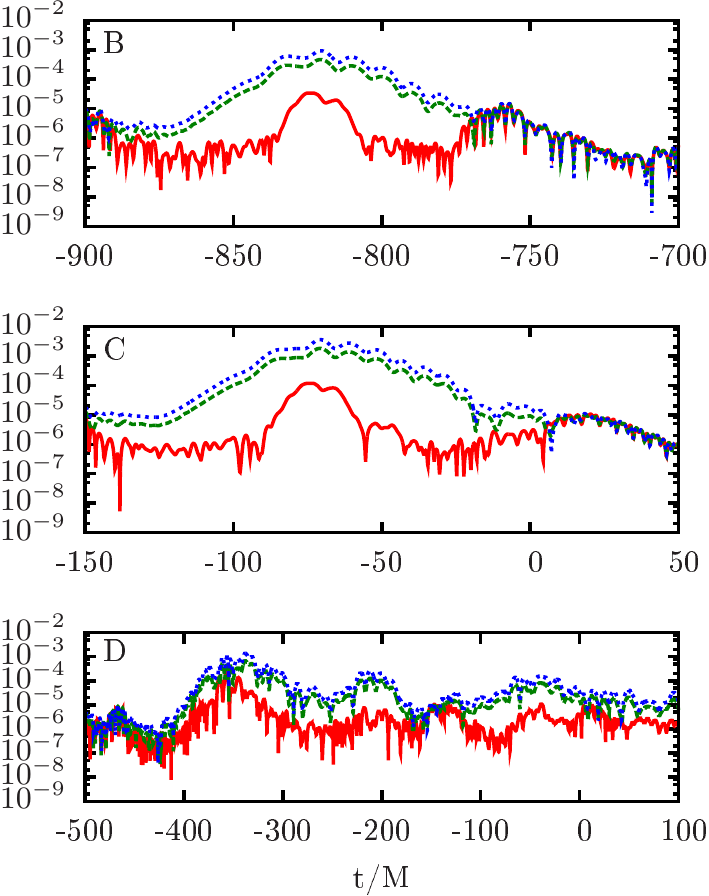}
  \caption{Same as in Fig.~\ref{fig:Phi00}  but for the mode (2,2).}
\label{fig:Phi22}
\end{figure}

The presence of a scalar field in \stg{} theories predicts dipole
energy losses as well as a new \gw{} polarization, sometimes called a
\emph{breathing mode}, given by the traceless Ricci tensor scalar,
$\Phi_{22} = -R_{l m l \bar m} =
-R_{ll}/2$~\cite{1973PhRvD...8.3308E,2010CQGra..27n5010A}.
Figures~\ref{fig:Phi00} and~\ref{fig:Phi22} show respectively the
modes (0,0) and (2,2) of $\widetilde\Phi_{22}$ in the Jordan frame. In
each panel, $b=0, 5$ and 10 are solid, dashed, and dotted lines,
respectively.  The time range in each panel was chosen to highlight
when $\phi$ interacts most strongly with the \bh{s}, with $t = 0\,M$
denoting merger time. The evident $b$-dependence of
$\widetilde\Phi_{22}$ can be explained by how $\widetilde\Phi_{22}$
conformally transforms: $\widetilde\Phi_{2,2} = A^{-2}\, [\Phi_{2,2} +
b\,D(\partial_t\phi,\dots)]$, with $D$ a function that vanishes when
$\partial_t\phi = 0$. Therefore, at times when $\phi$ undergoes
significant evolution, $\widetilde\Phi_{22}$ depends linearly on $b$,
as observed in Figs.~\ref{fig:Phi00} and ~\ref{fig:Phi22}. Overlaps of
$\widetilde\Phi_{22}$ with different $b$ in cases B and C correspond
to moments when $\partial_t\phi \approx 0$ at the wave extraction
location.  Then, as with $\Psi_4$, one has $\widetilde \Phi_{22}
\approx \Phi_{22}$. For case D, this does not happen, i.e.~$\widetilde
\Phi_{22}$ with different $b$ values do not overlap because the
potential induces longer lived dynamics in $\phi$, persisting long
after the binary has merged.

This study is a first step towards investigating detectable
observational signatures in the \gw{} emission from the late inspiral
and merger of a \bbh{} in \stg{} gravity.  Our results
supports the view that, in order to ``defeat'' the no-hair constraint
of \bh{s}, and thus trigger detectable effects on the gravitational
radiation, one needs a mechanism to excite scalar dynamics. 
In our study, we achieved this with inhomogeneities in the
form of a scalar field bubble surrounding the \bh{s}. The effect on the
binary was dramatic for the chosen parameters, mostly due to the
accretion of scalar field by the merging \bh{s}.  In a subsequent
study, we will explore a broader range of parameters, particularly
those compatible with the parameterized post-Newtonian bounds. We will
also consider a situation with a cosmological scalar background, and
the inspiral and merger of mixed binary systems.

We thank Cliff Will for fruitful discussions. Work supported by NSF
grants 0653443, 0855892, 0914553, 0941417, 0903973, 0955825, 1114374
and NASA grant NNX11AI49G, under sub-award 00001944. Computations at
Teragrid TG-MCA08X009 and Georgia Tech FoRCE cluster. RH gratefully
acknowledges support by the Natural Sciences and Engineering Council
of Canada.

%\bibliography{refs}

\begin{thebibliography}{26}
\expandafter\ifx\csname natexlab\endcsname\relax\def\natexlab#1{#1}\fi
\expandafter\ifx\csname bibnamefont\endcsname\relax
  \def\bibnamefont#1{#1}\fi
\expandafter\ifx\csname bibfnamefont\endcsname\relax
  \def\bibfnamefont#1{#1}\fi
\expandafter\ifx\csname citenamefont\endcsname\relax
  \def\citenamefont#1{#1}\fi
\expandafter\ifx\csname url\endcsname\relax
  \def\url#1{\texttt{#1}}\fi
\expandafter\ifx\csname urlprefix\endcsname\relax\def\urlprefix{URL }\fi
\providecommand{\bibinfo}[2]{#2}
\providecommand{\eprint}[2][]{\url{#2}}

\bibitem[{\citenamefont{Will}(2006)}]{will:2006:cbg}
\bibinfo{author}{\bibfnamefont{C.~M.} \bibnamefont{Will}},
  \bibinfo{journal}{Living Reviews in Relativity} \textbf{\bibinfo{volume}{9}}
  (\bibinfo{year}{2006}).

\bibitem[{\citenamefont{Taylor et~al.}(1992)\citenamefont{Taylor, Wolszczan,
  Damour, and Weisberg}}]{126}
\bibinfo{author}{\bibfnamefont{J.}~\bibnamefont{Taylor}},
  \bibinfo{author}{\bibfnamefont{A.}~\bibnamefont{Wolszczan}},
  \bibinfo{author}{\bibfnamefont{T.}~\bibnamefont{Damour}}, \bibnamefont{and}
  \bibinfo{author}{\bibfnamefont{J.}~\bibnamefont{Weisberg}},
  \bibinfo{journal}{Nature} \textbf{\bibinfo{volume}{355}},
  \bibinfo{pages}{132} (\bibinfo{year}{1992}).

\bibitem[{\citenamefont{{Damour} and
  {Esposito-Farese}}(1992)}]{1992CQGra...9.2093D}
\bibinfo{author}{\bibfnamefont{T.}~\bibnamefont{{Damour}}} \bibnamefont{and}
  \bibinfo{author}{\bibfnamefont{G.}~\bibnamefont{{Esposito-Farese}}},
  \bibinfo{journal}{Classical and Quantum Gravity}
  \textbf{\bibinfo{volume}{9}}, \bibinfo{pages}{2093} (\bibinfo{year}{1992}).

\bibitem[{\citenamefont{Damour and Esposito-Farese}(1993)}]{Damour:1993hw}
\bibinfo{author}{\bibfnamefont{T.}~\bibnamefont{Damour}} \bibnamefont{and}
  \bibinfo{author}{\bibfnamefont{G.}~\bibnamefont{Esposito-Farese}},
  \bibinfo{journal}{Phys. Rev. Lett.} \textbf{\bibinfo{volume}{70}},
  \bibinfo{pages}{2220} (\bibinfo{year}{1993}).

\bibitem[{\citenamefont{Peebles and Ratra}(2003)}]{11496}
\bibinfo{author}{\bibfnamefont{P.}~\bibnamefont{Peebles}} \bibnamefont{and}
  \bibinfo{author}{\bibfnamefont{B.}~\bibnamefont{Ratra}},
  \bibinfo{journal}{Rev. Mod. Phys.} \textbf{\bibinfo{volume}{75}},
  \bibinfo{pages}{559} (\bibinfo{year}{2003}).

\bibitem[{\citenamefont{Brans and Dicke}(1961)}]{Brans:1961sx}
\bibinfo{author}{\bibfnamefont{C.}~\bibnamefont{Brans}} \bibnamefont{and}
  \bibinfo{author}{\bibfnamefont{R.~H.} \bibnamefont{Dicke}},
  \bibinfo{journal}{Phys. Rev.} \textbf{\bibinfo{volume}{124}},
  \bibinfo{pages}{925} (\bibinfo{year}{1961}).

\bibitem[{\citenamefont{{Will} and {Zaglauer}}(1989)}]{1989ApJ...346..366W}
\bibinfo{author}{\bibfnamefont{C.~M.} \bibnamefont{{Will}}} \bibnamefont{and}
  \bibinfo{author}{\bibfnamefont{H.~W.} \bibnamefont{{Zaglauer}}},
  \bibinfo{journal}{\apj} \textbf{\bibinfo{volume}{346}}, \bibinfo{pages}{366}
  (\bibinfo{year}{1989}).

\bibitem[{\citenamefont{{Scharre} and {Will}}(2002)}]{2002PhRvD..65d2002S}
\bibinfo{author}{\bibfnamefont{P.~D.} \bibnamefont{{Scharre}}}
  \bibnamefont{and} \bibinfo{author}{\bibfnamefont{C.~M.}
  \bibnamefont{{Will}}}, \bibinfo{journal}{\prd} \textbf{\bibinfo{volume}{65}},
  \bibinfo{pages}{042002} (\bibinfo{year}{2002}), \eprint{arXiv:gr-qc/0109044}.

\bibitem[{\citenamefont{{Yagi} and {Tanaka}}(2010)}]{2010PhRvD..81f4008Y}
\bibinfo{author}{\bibfnamefont{K.}~\bibnamefont{{Yagi}}} \bibnamefont{and}
  \bibinfo{author}{\bibfnamefont{T.}~\bibnamefont{{Tanaka}}},
  \bibinfo{journal}{\prd} \textbf{\bibinfo{volume}{81}},
  \bibinfo{pages}{064008} (\bibinfo{year}{2010}), \eprint{0906.4269}.

\bibitem[{\citenamefont{{Stavridis} and {Will}}(2009)}]{2009PhRvD..80d4002S}
\bibinfo{author}{\bibfnamefont{A.}~\bibnamefont{{Stavridis}}} \bibnamefont{and}
  \bibinfo{author}{\bibfnamefont{C.~M.} \bibnamefont{{Will}}},
  \bibinfo{journal}{\prd} \textbf{\bibinfo{volume}{80}},
  \bibinfo{pages}{044002} (\bibinfo{year}{2009}), \eprint{0906.3602}.

\bibitem[{\citenamefont{{Berti} et~al.}(2005)\citenamefont{{Berti}, {Buonanno},
  and {Will}}}]{2005CQGra..22S.943B}
\bibinfo{author}{\bibfnamefont{E.}~\bibnamefont{{Berti}}},
  \bibinfo{author}{\bibfnamefont{A.}~\bibnamefont{{Buonanno}}},
  \bibnamefont{and} \bibinfo{author}{\bibfnamefont{C.~M.}
  \bibnamefont{{Will}}}, \bibinfo{journal}{Classical and Quantum Gravity}
  \textbf{\bibinfo{volume}{22}}, \bibinfo{pages}{943} (\bibinfo{year}{2005}),
  \eprint{arXiv:gr-qc/0504017}.

\bibitem[{\citenamefont{Will}(1994)}]{Will:1994fb}
\bibinfo{author}{\bibfnamefont{C.~M.} \bibnamefont{Will}},
  \bibinfo{journal}{Phys. Rev.} \textbf{\bibinfo{volume}{D50}},
  \bibinfo{pages}{6058} (\bibinfo{year}{1994}), \eprint{gr-qc/9406022}.

\bibitem[{\citenamefont{{Will} and {Yunes}}(2004)}]{2004CQGra..21.4367W}
\bibinfo{author}{\bibfnamefont{C.~M.} \bibnamefont{{Will}}} \bibnamefont{and}
  \bibinfo{author}{\bibfnamefont{N.}~\bibnamefont{{Yunes}}},
  \bibinfo{journal}{Classical and Quantum Gravity}
  \textbf{\bibinfo{volume}{21}}, \bibinfo{pages}{4367} (\bibinfo{year}{2004}),
  \eprint{arXiv:gr-qc/0403100}.

\bibitem[{\citenamefont{{Yagi} et~al.}(2011)\citenamefont{{Yagi}, {Stein},
  {Yunes}, and {Tanaka}}}]{2011arXiv1110.5950Y}
\bibinfo{author}{\bibfnamefont{K.}~\bibnamefont{{Yagi}}},
  \bibinfo{author}{\bibfnamefont{L.~C.} \bibnamefont{{Stein}}},
  \bibinfo{author}{\bibfnamefont{N.}~\bibnamefont{{Yunes}}}, \bibnamefont{and}
  \bibinfo{author}{\bibfnamefont{T.}~\bibnamefont{{Tanaka}}},
  \bibinfo{journal}{ArXiv e-prints}  (\bibinfo{year}{2011}),
  \eprint{1110.5950}.

\bibitem[{\citenamefont{{Gair} and {Yunes}}(2011)}]{2011PhRvD..84f4016G}
\bibinfo{author}{\bibfnamefont{J.}~\bibnamefont{{Gair}}} \bibnamefont{and}
  \bibinfo{author}{\bibfnamefont{N.}~\bibnamefont{{Yunes}}},
  \bibinfo{journal}{\prd} \textbf{\bibinfo{volume}{84}}, \bibinfo{eid}{064016}
  (\bibinfo{year}{2011}), \eprint{1106.6313}.

\bibitem[{\citenamefont{{Yunes} et~al.}(2011)\citenamefont{{Yunes}, {Pani}, and
  {Cardoso}}}]{2011arXiv1112.3351Y}
\bibinfo{author}{\bibfnamefont{N.}~\bibnamefont{{Yunes}}},
  \bibinfo{author}{\bibfnamefont{P.}~\bibnamefont{{Pani}}}, \bibnamefont{and}
  \bibinfo{author}{\bibfnamefont{V.}~\bibnamefont{{Cardoso}}},
  \bibinfo{journal}{ArXiv e-prints}  (\bibinfo{year}{2011}),
  \eprint{1112.3351}.

\bibitem[{\citenamefont{{Horbatsch} and {Burgess}}(2011)}]{2011arXiv1111.4009H}
\bibinfo{author}{\bibfnamefont{M.~W.} \bibnamefont{{Horbatsch}}}
  \bibnamefont{and} \bibinfo{author}{\bibfnamefont{C.~P.}
  \bibnamefont{{Burgess}}}, \bibinfo{journal}{ArXiv e-prints}
  (\bibinfo{year}{2011}), \eprint{1111.4009}.

\bibitem[{\citenamefont{{Jacobson}}(1999)}]{1999PhRvL..83.2699J}
\bibinfo{author}{\bibfnamefont{T.}~\bibnamefont{{Jacobson}}},
  \bibinfo{journal}{Physical Review Letters} \textbf{\bibinfo{volume}{83}},
  \bibinfo{pages}{2699} (\bibinfo{year}{1999}),
  \eprint{arXiv:astro-ph/9905303}.

\bibitem[{\citenamefont{{Yunes} and {Stein}}(2011)}]{2011PhRvD..83j4002Y}
\bibinfo{author}{\bibfnamefont{N.}~\bibnamefont{{Yunes}}} \bibnamefont{and}
  \bibinfo{author}{\bibfnamefont{L.~C.} \bibnamefont{{Stein}}},
  \bibinfo{journal}{\prd} \textbf{\bibinfo{volume}{83}}, \bibinfo{eid}{104002}
  (\bibinfo{year}{2011}), \eprint{1101.2921}.

\bibitem[{\citenamefont{{Yunes} and {Pretorius}}(2009)}]{2009PhRvD..79h4043Y}
\bibinfo{author}{\bibfnamefont{N.}~\bibnamefont{{Yunes}}} \bibnamefont{and}
  \bibinfo{author}{\bibfnamefont{F.}~\bibnamefont{{Pretorius}}},
  \bibinfo{journal}{\prd} \textbf{\bibinfo{volume}{79}}, \bibinfo{eid}{084043}
  (\bibinfo{year}{2009}), \eprint{0902.4669}.

\bibitem[{\citenamefont{{Alexander} and {Yunes}}(2009)}]{2009PhR...480....1A}
\bibinfo{author}{\bibfnamefont{S.}~\bibnamefont{{Alexander}}} \bibnamefont{and}
  \bibinfo{author}{\bibfnamefont{N.}~\bibnamefont{{Yunes}}},
  \bibinfo{journal}{Physics Reports} \textbf{\bibinfo{volume}{480}},
  \bibinfo{pages}{1} (\bibinfo{year}{2009}), \eprint{0907.2562}.

\bibitem[{\citenamefont{Gelmini et~al.}(1989)\citenamefont{Gelmini, Gleiser,
  and Kolb}}]{PhysRevD.39.1558}
\bibinfo{author}{\bibfnamefont{G.~B.} \bibnamefont{Gelmini}},
  \bibinfo{author}{\bibfnamefont{M.}~\bibnamefont{Gleiser}}, \bibnamefont{and}
  \bibinfo{author}{\bibfnamefont{E.~W.} \bibnamefont{Kolb}},
  \bibinfo{journal}{Phys. Rev. D} \textbf{\bibinfo{volume}{39}},
  \bibinfo{pages}{1558} (\bibinfo{year}{1989}),
  \urlprefix\url{http://link.aps.org/doi/10.1103/PhysRevD.39.1558}.

\bibitem[{\citenamefont{Bode et~al.}(2010)\citenamefont{Bode, Haas, Bogdanovic,
  Laguna, and Shoemaker}}]{Bode:2009mt}
\bibinfo{author}{\bibfnamefont{T.}~\bibnamefont{Bode}},
  \bibinfo{author}{\bibfnamefont{R.}~\bibnamefont{Haas}},
  \bibinfo{author}{\bibfnamefont{T.}~\bibnamefont{Bogdanovic}},
  \bibinfo{author}{\bibfnamefont{P.}~\bibnamefont{Laguna}}, \bibnamefont{and}
  \bibinfo{author}{\bibfnamefont{D.}~\bibnamefont{Shoemaker}},
  \bibinfo{journal}{Astrophys. J.} \textbf{\bibinfo{volume}{715}},
  \bibinfo{pages}{1117} (\bibinfo{year}{2010}), \eprint{0912.0087}.

\bibitem[{\citenamefont{{L{\"o}ffler} et~al.}(2011)\citenamefont{{L{\"o}ffler},
  {Faber}, {Bentivegna}, {Bode}, {Diener}, {Haas}, {Hinder}, {Mundim}, {Ott},
  {Schnetter} et~al.}}]{2011arXiv1111.3344L}
\bibinfo{author}{\bibfnamefont{F.}~\bibnamefont{{L{\"o}ffler}}},
  \bibinfo{author}{\bibfnamefont{J.}~\bibnamefont{{Faber}}},
  \bibinfo{author}{\bibfnamefont{E.}~\bibnamefont{{Bentivegna}}},
  \bibinfo{author}{\bibfnamefont{T.}~\bibnamefont{{Bode}}},
  \bibinfo{author}{\bibfnamefont{P.}~\bibnamefont{{Diener}}},
  \bibinfo{author}{\bibfnamefont{R.}~\bibnamefont{{Haas}}},
  \bibinfo{author}{\bibfnamefont{I.}~\bibnamefont{{Hinder}}},
  \bibinfo{author}{\bibfnamefont{B.~C.} \bibnamefont{{Mundim}}},
  \bibinfo{author}{\bibfnamefont{C.~D.} \bibnamefont{{Ott}}},
  \bibinfo{author}{\bibfnamefont{E.}~\bibnamefont{{Schnetter}}},
  \bibnamefont{et~al.}, \bibinfo{journal}{ArXiv e-prints}
  (\bibinfo{year}{2011}), \eprint{1111.3344}.

\bibitem[{\citenamefont{{Eardley} et~al.}(1973)\citenamefont{{Eardley}, {Lee},
  and {Lightman}}}]{1973PhRvD...8.3308E}
\bibinfo{author}{\bibfnamefont{D.~M.} \bibnamefont{{Eardley}}},
  \bibinfo{author}{\bibfnamefont{D.~L.} \bibnamefont{{Lee}}}, \bibnamefont{and}
  \bibinfo{author}{\bibfnamefont{A.~P.} \bibnamefont{{Lightman}}},
  \bibinfo{journal}{\prd} \textbf{\bibinfo{volume}{8}}, \bibinfo{pages}{3308}
  (\bibinfo{year}{1973}).

\bibitem[{\citenamefont{{Alves} et~al.}(2010)\citenamefont{{Alves}, {Miranda},
  and {de Araujo}}}]{2010CQGra..27n5010A}
\bibinfo{author}{\bibfnamefont{M.~E.~S.} \bibnamefont{{Alves}}},
  \bibinfo{author}{\bibfnamefont{O.~D.} \bibnamefont{{Miranda}}},
  \bibnamefont{and} \bibinfo{author}{\bibfnamefont{J.~C.~N.} \bibnamefont{{de
  Araujo}}}, \bibinfo{journal}{Classical and Quantum Gravity}
  \textbf{\bibinfo{volume}{27}}, \bibinfo{pages}{145010}
  (\bibinfo{year}{2010}), \eprint{1004.5580}.

\end{thebibliography}

\end{document}